\documentstyle[preprint, aps, epsf, psfig]{revtex}

\begin{document}
\draft
\preprint{}
\title{Surface-Directed Spinodal Decomposition in Binary Fluid Mixtures}
\author{Sorin Bastea$^{1,}$\cite{sb}, Sanjay Puri$^2$ and Joel L. Lebowitz$^1$}
\address{$^1$Department of Mathematics and Physics,
Rutgers University, New Brunswick, NJ 08903, USA\\
$^2$School of Physical Sciences, Jawaharlal
Nehru University, New Delhi 110067, INDIA}
\maketitle
\begin{abstract}
We consider the phase separation of binary fluids in contact with a
surface which is preferentially wetted by one of the components of the
mixture. We review the results available for this problem and present 
new numerical results obtained using a mesoscopic-level simulation technique
for the 3-dimensional problem.\\
\end{abstract}
\widetext

\section{Introduction}

There has been much interest in the phase-separation dynamics of
homogeneous binary mixtures, which have been rendered thermodynamically
unstable by a rapid quench below the coexistence curve. The time
evolution
of pure bulk mixtures in which the evolving system coarsens into domains
rich in either of the components is now reasonably well-understood. 
These
domains are characterized for late times by a single length scale $L(t)
\sim t^{\phi}$, where $t$ is the time and the growth exponent $\phi$
depends upon the system considered, e.g., whether or not the order
parameter is conserved, the relevance of hydrodynamic effects,
etc. \cite{rev}.

An experimentally important variation of this problem considers the role
of
surfaces with a preferential attraction for one of the components of the
mixture. The first experimental study of this problem is due to Jones
{\it
et al.} \cite{jon}, who considered unstable polymer mixtures of
polyethylene-propylene (PEP) and perdeuterated PEP (d-PEP) in a
thin-film
geometry. The surface energy of d-PEP is somewhat less than that of PEP
leading, in addition to bulk phase separation (spinodal decomposition),
to
a preferential deposition of d-PEP at any free surface.
Jones {\it et al.} studied laterally-averaged composition profiles as a
function of distance from the surface. The bulk is characterized by
randomly-oriented phase-separation profiles and the lateral averaging
procedure does not yield a systematic behavior. However, the surface
exhibits an enrichment layer in the preferred component, which is
followed by a depletion layer. This oscillatory profile is 
time-dependent and decays with a characteristic length to the 
bulk composition.

This experiment motivated many further investigations of this problem.  The
experimental techniques and results have been reviewed by Krausch
\cite{kra}, and the theoretical and numerical developments by Puri and
Frisch \cite{pur} and Binder \cite{pur}. To date, most numerical studies of
this problem have focused on the case of binary mixtures without
hydrodynamic effects, i.e., the growth of surface wetting layers and bulk
domains is governed by diffusive processes. However, many important
experiments in this area involve binary fluids in contact with a
surface. It is well-known that macroscopic matter and energy flows, i.e.,
hydrodynamic effects, drastically alter the nature of domain growth in the
bulk phase-separation problem. Therefore, it is reasonable to expect
important physical effects to result from hydrodynamic flows in the case of
surface-directed phase-separation also. To understand some of the issues
involved, we have undertaken a detailed numerical simulation of this
problem.  In particular we adapted mesoscopic models formulated to study
bulk spinodal decomposition in binary fluids to surface-directed spinodal
decomposition.

This paper is organized as follows. Section 2 reviews 
available experimental, analytical and numerical results for this 
problem. In Section 3, we describe our model and the numerical methods
used.  These involve an ``integration'' of the 
Vlasov-Boltzmann equations for the binary
mixture in contact with a surface. In Section 4, we present 
results obtained from our simulations. Finally, Section 5 is devoted to
a summary and discussion of the results.

\section{Summary of Available Results}

\subsection{Experimental Studies}

One of the earliest experiments on phase-separating binary fluids near a
surface is due to Guenoun {\it et al.} \cite{gue}, which considered
unstable mixtures of cyclohexane (C) and methanol (M) in contact with a
surface which preferred M. The surface rapidly developed a M-rich
layer,
followed by a bicontinuous domain structure. Guenoun {\it et al.} found
that
domain growth was characterized by a number of different length scales.
Thus, the wetting layer grew as $R_1(t) \sim t^{a}$ with $a \simeq 
0.56$. The domains adjacent to the wetting layer were anisotropic 
and were characterized
by perpendicular $(L_{\bot} (t) \sim t^{b}$ with $b \simeq 0.64$) and 
parallel $(L_{\|}(t) \sim t^{c}$ with $c \simeq 1$) scales. 

Wiltzius {\it et al.}
\cite{wil} considered critical fluid mixtures of polyisoprene (PI) and
PEP sealed between two quartz plates. They found that the structure
factor exhibited two peaks -
one corresponding to the usual bulk domain length scale $L_b(t) \sim t$
and the other corresponding to a fast length scale $L_s(t) \sim
t^{3/2}$.
Furthermore, they found that the dimensionality of domain growth
associated
with the fast length scale was $d=2$, suggesting that it resulted from a
rapid coarsening in the surface layer. The rapid surface growth was
interpreted 
as a prelude to the formation of a
complete wetting layer on the surface. In that case, there is no
inconsistency between their results and the earlier results of Guenoun
{\it et al.} \cite{gue}, which correspond to later times when a
complete wetting layer was already formed. Similar experiments were
also performed by Shi and collaborators \cite{bqs} on mixtures of
guaiacol
and glycerol-water confined in a thin-film geometry. 

Detailed studies of the morphologies which arise for 
phase-separating mixtures confined to 1- and 2-dimensional capillaries 
were performed by Tanaka and co-workers \cite{tan} on critical and 
off-critical mixtures of polyvinyl-methyl-ether (PVME) and water, and 
$\epsilon$-caprolactone oligomer (OCL) and styrene oligomer (OS).
In particular, they clarified conditions under which the equilibrium
state is completely wet (i.e., only the preferred phase is in contact
with the surface) or partially wet (i.e., both phases are
in contact with the surface). Tanaka's group
did not observe the fast growth reported by Wiltzius and co-workers
\cite{wil,bqs}, possibly because the quench depth in their
experiments 
was too large. Once the wetting layer is formed, they found that its
thickness grows linearly in time, i.e., $R_1(t) \sim t$, in
disagreement with the experiments of Guenoun {\it et al.} \cite{gue}.
In most of their experiments,
the wetting layer is finally destabilized by a Rayleigh instability and
the system crosses over to a partially wet morphology.

\subsection{Analytical Arguments}

The equilibrium behavior of immiscible binary fluids in contact with a
substrate was examined long ago by Young \cite{you}. Let $\gamma_A$ and 
$\gamma_B$ be the surface energies
per unit area for the fluids A and B in contact with the substrate (say,
$\gamma_B > \gamma_A$); and let $\sigma$ be the surface tension between
fluids A and B. Then, the contact angle $\theta$ between A and the
surface is given by $\sigma \cos \theta = \gamma_B - \gamma_A$.
This equation has no solution when $(\gamma_B - \gamma_A)/\sigma > 1$, 
which corresponds to a situation where the preferred fluid
(A) completely wets the substrate. The effects of geometry and
composition can 
also be included \cite{tan,liu,cah}.

The nonequilibrium problem we consider is a homogeneous critical binary
mixture 
(at high temperature) in contact with a surface which has a preference
for one 
of the components of the mixture. At time $t=0$ the system is quenched
below its 
critical temperature and becomes
unstable to phase separation. We are interested in the dynamics of
approach to
the equilibrium morphology, which will consist
of either partially wet  (PW) or completely wet (CW) configurations. 
Typically, the surface is initially coated 
by the preferred component, which is then followed by the growth of the 
wetting layer \cite{tan}. We focus here on the wetting layer growth. 

As remarked 
by Siggia \cite{sig} the bicontinous morphology of critical or near
critical phase 
separating binary fluids 
consists essentially of interpenetrating ``tubes''. When a tube of the 
preferred 
phase establishes contact with the surface layer the curvature induced 
pressure gradient $\sigma/L^2$ 
leads to a flow of material from the tube to the surface. The material
flux per tube can 
be estimated for example from Poiseuille law to be $(\sigma/\eta) L^2$
\cite{sig}. 
Then $S(dR_1/dt) \sim (\sigma/\eta) L(t)^2 \times (S/L(t)^2)$, where $S$
is
the surface area and $S/L(t)^2$ is the number of tubes. Thus, 
$R_1(t) \sim (\sigma/\eta) t$ for the hydrodynamic problem, a result
which has been confirmed experimentally \cite{tan}. We believe
that the discrepancy between this result and the earlier experimental
work of 
Guenoun {\it et al.} \cite{gue} is due to the long-lived transient
growth laws
dependent upon the form of the surface potential \cite{pdb}. For the
diffusive case 
the chemical potential gradient 
between the bulk tube, $\mu \sim (\sigma/L)$ \cite{rev}, and the flat
tube portion 
at the surface, $\mu \sim 0$, induces a current $j \sim
(\sigma/L^2)$ and therefore a flux per tube $\sim\sigma$. 
The corresponding growth law is then $R_1(t) \sim \sigma^{1/3} t^{1/3}$.

The wetting layer grows until it reaches the equilibrium length
(dictated by the composition for a CW morphology), or is destabilized by
surface
fluctuations and goes over to the appropriate equilibrium PW morphology.
There is also a dynamical coupling of phase separation and
the growth of the wetting layer, which leads to the domains adjacent to
the
wetting layer being anisotropic with $L_{\bot}(t) < L_{\|}(t)$
\cite{gue,pur}. 

\subsection{Numerical Results}

One of the earliest numerical studies of the hydrodynamic problem is due
to
Keblinski {\it et al.} \cite{keb}, who performed molecular dynamics
(MD) simulations of binary fluids (AB) confined in a 2-dimensional
capillary (or a planar thin
film). One of the cases that they study is when the wall preferentially
attracts A, which is
analogous to the
experiments we have discussed earlier. In this case they observe a
``fast mode'' in the surface layer but with an exponential growth rather
than the power law growth.

Chen and Chakrabarti \cite{che} have studied phase
separation in 2-dimensional binary fluids near a surface through
numerical solutions
of the coarse-grained model H equations \cite{rev} in a semi-infinite
geometry. The model H equations consist of coupled dynamical equations
for the order parameter and the fluid velocity field. Chen and
Chakrabarti consider a surface with a long-range attraction for one of
the
components of the mixture and impose ``no-slip'' conditions on the
velocity field
at the surface. They find that the wetting-layer growth
crosses over from $R_1(t)
\sim t^{1/3}$ (characteristic of bulk diffusive growth in any dimension)
to $R_1(t) \sim t^{2/3}$ (characteristic of bulk hydrodynamic growth in
$d=2$). This crossover is associated with domains of the preferred
component establishing contact with the surface layer and
their subsequent rapid draining into the surface layer.

Another study of model H in a semi-infinite geometry is due to Tanaka
and
Araki \cite{ta}. These authors solved the model H equations numerically
in
$d=3$. They find that the wetting-layer thickness grows initially as
$R_1(t) \sim t^{1/3}$ (characteristic of diffusive growth) and then
crosses
over to the hydrodynamic regime with $R(t) \sim t$. They also study
characteristic
length scales in the layer parallel to the surface. Far from the
surface,
they find the expected bulk growth law $L_{\|}(t) \sim t$, while in the
vicinity of the surface they find a faster growth. However, it seems
difficult to unambiguously assign an exponent to this faster growth.
Furthermore, the time-regime of the ``fast mode'' is considerably later
than the time-scale of formation of the complete wetting layer. This
suggests to us that the ``fast growth'' observed by Tanaka and Araki
should
be identified with the anisotropic growth (with $L_{\bot} < L_{\|}$) of
domains in the vicinity of the wetting layer due to orientational
effects
of the wetting layer, rather than the ``fast mode'' of Wiltzius and
others
\cite{wil,bqs}. As we have discussed earlier, this fast mode is
associated with the coating dynamics which results in a complete wetting
layer.

Finally, we mention a MD study by Toxvaerd \cite{tox} who
investigated critical mixtures (AB) of particles interacting through
Lennard-Jones potentials. He focused on the morphologies which arise
for different wall-types, e.g., one wall attracts A whereas the other
wall
attracts B versus the case where both walls attract A and B equally,
etc. Using MD simulation he finds that the system evolves into a
layered
morphology, with the layer being parallel to the surface walls. We
believe
that these are metastable configurations which evolve exceedingly slowly
due to the low effective dimensionality $(d=1)$ of the system.

The paucity of detailed numerical results for binary fluids undergoing
phase
separation in contact with a wetting surface motivated us to 
undertake a mesoscopic-level simulation of this problem, through a
direct ``solution'' of the relevant Vlasov-Boltzmann equations.

\section{Description of Model}

The subtle interplay between diffusion and convection which occurs in
phase-separating fluids makes the modeling of these systems more
complicated than that of solids. The local conservation of linear
momentum
and energy, which is characteristic of fluids, and the associated
transport
of matter and energy on macroscopic scales, plays a crucial role during
phase segregation. Typically, the phase separation of binary fluids has
been modeled either (a) at the microscopic level, e.g., via MD
simulations,
or (b) at the macroscopic level via coarse-grained hydrodynamic
equations. An alternative to these approaches was introduced in
Ref. \cite{bl}.  The system studied was a binary mixture consisting of
$A$ and $B$ particles with short range repulsive interactions, modeled by hard spheres 
with equal mass $m$ and diameter $d$, and a long-range Gaussian repulsion between
the two components, $A$ and $B$.  The dynamics was described by coupled Vlasov-Boltzmann
kinetic equations:
\begin{equation}
\frac{\partial f_i}{\partial t}+{\bf v} \cdot \frac{\partial
f_i}{\partial {\bf r}}+\frac{{\bf F}_i}{m}\cdot\frac{\partial
f_i}{\partial {\bf v}}=J[f_i,f_1 + f_2]\hspace{0.5in}i=1,2
\end{equation}
where $f_i({\bf r},{\bf v},t)$ are the one-particle distribution functions,
${\bf F}_i({\bf r},t)=-\nabla V_i({\bf r})$, 
$V_i({\bf r})=\int V(|{\bf r}-{\bf r}\prime|)n_j({\bf
r}\prime)d{\bf r}\prime$ (Vlasov potential), 
$n_j({\bf r}\prime)=\int f_j({\bf r}\prime, {\bf v},
t)d{\bf v}$, with $i\neq j$, and $J[f,g]$ is
the Boltzmann collision operator for hard core interactions \cite{chapman1}.
The Boltzmann equation properly describes the dynamics of dilute gases, where 
the free flow of the particles is interrupted by localized binary collisions between 
particles that are uncorrelated. The Vlasov term takes into account the long-range 
interaction in the spirit of the mean-field approximation: each particle now moves 
between collisions in the background potential generated by all the other particles 
it is interacting with through the long-range potential $V(r)$.

The above mesoscopic representation in terms of one-particle distribution functions 
has the advantage that the relevant conservation laws are automatically satisfied, 
and it also provides a rigorous route to a macroscopic description
\cite{belm1}. Computationally, the method introduced in Ref. \cite{bl,sb1} to 
simulate the Vlasov-Boltzmann kinetics at the particle level,
i.e., coupling of the direct simulation Monte Carlo (DSMC) algorithm \cite{bird1}
for close-range collisions and the grid-weighting method 
for the long-range repulsions \cite{langdon1}, contains the essential physical 
ingredients of the Vlasov-Boltzmann equations, and it permits the study of much
bigger systems than those used in MD calculations.

In the present work, we modify the model of Ref. \cite{bl} to include
the presence of a preferred surface. 
One of the components of the binary mixture (say, A) interacts with 
the surface located at $z=0$ through an attractive
potential $W(z)$, which decays as 
$z^{-3}$ at large distances, i.e., $W(z)=-W_0$ if 
$z\leq\ r_0$ and $-W_0(r_0/z)^3$ otherwise, where 
$W_0>0$. This interaction potential corresponds to the case of
non-retarded van der Waals interactions in $d=3$ \cite{hm}.
The wall is diffusive \cite{cc}, i.e., particles ``hitting'' the
wall are absorbed and 
re-emitted isotropically with a velocity drawn from a Maxwellian
distribution with the 
temperature of the wall, $T_W$. The other wall along the $z$ direction 
is purely reflective (no preferred attraction), which allows us to run 
the simulations for longer times than if the set-up was symmetric. 
We performed simulations with equal fractions of the components 
at fixed temperature, $T/T_c=0.6$, $T_W = T$, where $T_c$ is the bulk
mean-field critical temperature of the
system, using the velocity-rescaling technique introduced by Berendsen
{\it et al.} \cite{berend}.  Below $T_c$ the bulk fluid segregates into
an
A rich and a B rich components, denoted by 1 and 2 respectively. 
The parameter varied was the strength of the 
wall-particle interaction $W_0$.

As discussed in subsection 2.2, the wall-wetting morphology, i.e.,
completely wet (CW) versus partially wet (PW), is determined by the ratio
$\sigma/(\gamma_2-\gamma_1)$, where $\sigma$ is the surface tension between
the two fluid phases 1 and 2; and $\gamma_1$, $\gamma_2$ are the
corresponding wall-fluid surface tensions. Taking into account that A and B
are partially miscible the surface tension parameters can be estimated as
follows. Consider the A-rich phase (1) with average total particle density
$n_0$ and average individual densities $n^0_{1A}$ and $n^0_{1B}$,
$n_0=n^0_{1A}+n^0_{1B}$.  The wall-fluid surface energy is then
\begin{equation}
\gamma_1=\int_0^\infty dz n_{1A}(z) W(z)
\end{equation}
If we neglect the
inter-particle spatial correlations, 
which is appropriate in our model if the long-range repulsion between
the two components is sufficiently weak, we can write
$n_{1A}(z)=n^0_{1A} \exp[-\beta W(z)]$. With this 
assumption and using the fact that the phases are symmetric, we obtain
the expression
\begin{equation}
\gamma_2-\gamma_1 = k_B T n_0 \phi_0 r_0 H(\beta W_0) ,
\end{equation} 
where $\pm \phi_0$ is the average order parameter in the two phases,
$\phi_0 = (n^0_{1A} - n^0_{1B})/n_0$, and the function $H(x)$ depends
on the wall-interaction potential. For our choice of wall-particle
interaction 
we have 
$$
H(x)=x [\exp(x)+(1/2) \int_0^1 dy \exp(xy^\frac{3}{2})]
$$

The surface tension $\sigma$ between the two phases 1 and 2 is related to
the profile of the planar interface separating the equilibrium phases
\cite{rw}. For our system this can be written as 
\begin{equation}
\sigma = m\int dz\{(d[n\phi]/dz)^2-(dn/dz)^2\}
\end{equation}
\cite{rw,bs}, where $n(z) \phi(z) = n_A(z)-n_B(z)$,
$n(z) = n_A (z)+n_B (z)$, and $m=(1/12)\int d{\bf r} V(r)r^2$, with
$V(r)$ 
the long-range repulsive interaction between the two species.  (NB. Here
$z$ stands for the distance from the center of a planar interface
separating
the two bulk phases.)   As remarked in \cite{bl} and is well-known for
these
systems \cite{rw}, $\phi(z)$ is well represented as $\phi_0 \tanh
(z/2\xi)$, where $\xi$ is a correlation length that characterizes the
interface thickness.  Furthermore, the total density profile $n(z)$ is
well
characterized as $n_0 [1-\delta \mbox{sech} (z/2\xi)]$, where $\phi_0$
and
$n_0$ are the values of the order parameter and density far from the
interface.  With these considerations, the surface tension between
phases 1
and 2 can be computed as:
\begin{equation}
\sigma=\frac{k_B T_c n_0 G(\phi_0,\delta)}{4\gamma^2\xi} .
\end{equation}

Here $\gamma^{-1}$ is the range of the inter-species potential
$V(r)=\alpha\gamma^3U(\gamma r)$, where, as in \cite{bl}, we use
$U(x) = \pi^{-\frac{3}{2}} \exp (-x^2)$ (note that $k_BT_c=n_0\alpha/2$
\cite{bl}). 
The function $G(\phi_0,\delta)$ has a simple algebraic form, 
$$
G(\phi_0,\delta)=(2/3) \phi_0^2 - (1/3)\delta^2 +
(7/15)\delta^2\phi_0^2 - 
(\pi/4)\delta\phi_0^2
$$ 
Therefore, we obtain the desired ratio for Young's condition as
\begin{equation}
\frac{\sigma}{\gamma_2-\gamma_1} = \frac{T_c}{T} \frac{G(\phi_0,\delta)}
{\phi_0} \frac{1}{4 \gamma^2 \xi r_0 H(\beta W_0)} .
\end{equation}
The physical quantities $\xi$, $\delta$ and 
$\phi_0$ have been obtained 
in simulations of the interface profile at $T/T_c=0.6$ \cite{bl} as
$\xi\simeq 1.5\gamma^{-1}$, $\phi_0\simeq 0.8$, $\delta\simeq 0.2$, 
and we set $r_0=\gamma^{-1}$.
Recall that we vary the surface potential strength $W_0$, and keep
other parameters fixed as specified above. We then estimate
Young's condition as corresponding to $\beta W^Y_0\simeq 0.071$. 
We have performed simulations with $W_0/W^Y_0$ ranging from $0.67$ to 
about $5$. The size of the system was $60\times 60\times 120$ in units
of the potential range $\gamma^{-1}$, and the number of particles used 
was approximately $N = 2.5\times 10^6$. 
For each value of $W_0$ we averaged the results of 12-15 independent
runs, wherever statistical averaging was required. In the figures presented 
below the unit of 
length is $\gamma^{-1}$ and the unit of time is the mean-free time between 
collisions $\tau=\lambda/c$, $\lambda=(2^{\frac{1}{2}}\pi nd^2)^{-1}$, 
$c=(2k_BT/m)^{\frac{1}{2}}$, where $n$ is the overall particle density, $d$ 
is the hard sphere diameter and $T$ is the temperature.
\section{Numerical Results}

As discussed in the previous section, $W_0/W^Y_0<1$ and $W_0/W^Y_0>1$ 
should correspond to the PW and CW 
cases, respectively. In our subsequent discussion, we will refer to
$W_0/W^Y_0<1$ as the weak-field case, and $W_0/W^Y_0>1$ as the
strong-field case. Of course, we should stress 
that there are additional entropic effects, which have not been
accounted for in our calculation. In general, this would raise the
critical surface field for transition from PW to CW morphologies.

Fig. 1 shows 3-dimensional snapshots of the evolution for 
$W_0/W^Y_0=0.67$ at times $t=60, 120$ and $180$. The wall is 
located at $z=0$ (extreme right) and preferentially attracts A, though
it is not completely wetted by A. The enrichment layer (in A) at the
surface is followed by a depletion layer in A; and this layered
structure
deforms continuously into the bulk, as has been seen in various earlier
studies for both the diffusive \cite{pb}, \cite{pur} and hydrodynamic
 \cite{ta} cases. Fig. 2 shows laterally-averaged profiles
$\phi_{av} (z,t)$ vs. $z$ (depth from the surface) for the evolution 
depicted in Fig. 1
at times $t=50, 100, 200$ and $275$. These profiles are obtained by
averaging the order parameter profiles in the direction parallel to the
surface -- analogous to the corresponding experimental
situation \cite{kra}. There is a systematic profile at
the surface, which decays to zero (due to isotropic phase separation) in
the bulk. The systematic surface profile propagates into the bulk with
the passage of time. Furthermore, the degree of enrichment diminishes as
isotropic phase separation in the bulk destroys the layered structure at
the surface.

To characterize the morphology of the surface layer, Fig. 3 plots the first
and second zeros of the laterally-averaged profiles as a function of
time. After an initial transient regime, the position of the first and
second zeros grow approximately linearly in time. The linear growth of the
first zero results from hydrodynamic draining of the preferred material to
the surface through bulk tubes which make contact with the surface
layer. This is in accordance with the observation of Tanaka and Araki
\cite{ta}, and the mechanism for this was discussed in subsection 2.2.
Additionally, it is reasonable to expect that the overall composition of
the first and second layers should be comparable with the average
composition. Thus, we expect the second zero, $R_2(t)$, to exhibit the same
scaling behavior as the first zero, $R_1(t)$.

Next we consider the evolution for the (very) strong-field case, where
the surface is completely wetted by the preferred component. Fig. 4
shows 3-dimensional evolution pictures for the case $W_0/W^Y_0=4$
at $t=60, 120, 180$. Notice the
perfectly layered structure at the surface. Figs. 1 and 4 should be
compared with analogous pictures for the diffusive problem \cite{pur}.
Fig. 5 shows the corresponding laterally-averaged profiles at $t=50,
100, 200$ and 275. The broad features 
are the same as in Fig. 2, but the level of enrichment (depletion)
of A in the surface layer (next-to-surface layer) is much higher. 
This layered
structure evolves more slowly in time because the bulk domains have not
established contact with the surface layer on the time-scales of our
simulation. Thus, growth of the wetting layer occurs only through
diffusive transport of A from the bulk through the depletion layer in A.
There are two regimes for diffusive growth \cite{pdb}. 
In the first regime, the attractive force due to the surface 
potential gives a potential-dependent
growth law. In the asymptotic regime, the chemical potential gradient
between the domains in the bulk and the flat wetting layer gives rise
to an asymptotic growth law $R_1(t) \sim t^{1/3}$, provided that the
preferred component is not the majority component. The crossover time
between the first and second regimes depends on the strength and range
of
the surface potential \cite{pdb}. In the present case, we have a
long-ranged surface potential $(V(z) \sim z^{-3})$. The corresponding
exponent for the potential-dependent growth regime is $\phi=0.2$ 
(in general, $\phi={1 \over n+2}$ for $V(z) \sim z^{-n}$ \cite{pdb}); 
and the asymptotic exponent for diffusive growth is $\phi = 1/3$. 

The growth of the location of the first and second zeros of the
laterally-averaged profile for this case is shown as a function of
$t^{1/3}$ in Fig. 6.  This plot is consistent with growth driven by the
chemical potential gradient between curved domains in the bulk and the flat
wetting layer.  As we have remarked earlier, the strong layering inhibits
the operation of draining modes to the surface layer, which would give rise
to the expected asymptotic behavior $R_1(t) \sim t$. We also show in Fig. 7
the growth of the first zero of the laterally-averaged profiles for all
cases that we studied.  The sharpness of the transition between the growth
in the weak and strong field cases is remarkable and is in very good
agreement with our analysis of the partial and complete wetting
morphologies.

\section{Discussion}

Our mesoscopic-level modeling is in terms of coupled Vlasov-Boltzmann
equations for 
binary hard-sphere mixtures with additional long range interactions. The
major
advantage of our modeling is that it enables the study of much larger
systems than those accessible in MD simulations. At the same time, we
are still able to identify scales and parameters in terms of
microscopic quantities -- in contrast to modeling via coarse-grained
partial differential equations.

The present work has focused on the morphology and temporal evolution of
surface-directed spinodal decomposition waves in critical binary fluids.
We
have considered parameter values where the surface is either partially
wet
or completely wet in equilibrium. It is not our argument that the
asymptotic behavior of these two cases is different. Rather, we would
like
to stress the appearance of long-lived transient regimes which are
critically dependent on the morphology.  These should be of relevance in
the interpretation of experiments.  A confusing range of exponents have
been reported in various studies and we believe that the present work
helps
systematize these exponents.

(a) Let us first focus on the diffusive problem. These 
are relevant for binary fluids at early times.
Furthermore, if the composition of the binary fluid is such that the
domain morphology is not continuous, domain growth again proceeds
through diffusive processes \cite{rev}. For cases where the preferred
component is the minority component, the wetting layer exhibits a
potential-dependent growth law, $R_1(t) \sim t^{\frac{1}{n+2}}$ for
$V(z) \sim z^{-n}$ \cite{pdb}, which crosses over to a universal
growth law, $R_1(t) \sim t^{1/3}$. The crossover time depends on the
surface field strength and the wetting-layer morphology. For cases
where the preferred component is the majority component, a
potential-dependent growth law applies for all time \cite{pdb}. \\
(b) Let us next consider the hydrodynamic case. For strongly
off-critical compositions, hydrodynamic growth modes are inactive
because of the discontinuous domain morphology. As we have stated
earlier, the results quoted in (a) apply for this case. For bicontinuous
bulk morphologies, wetting-layer growth can be characterized as
follows. There is an early-time growth (``fast mode'', with $L_s (t)
\sim
t^{3/2}$) associated with the formation of a coating layer
\cite{wil,tan}.
Subsequently, we expect the wetting layer to exhibit the diffusive
behavior outlined in (a) above. The asymptotic regime ($R_1(t) \sim t$)
is accessed when the bulk establishes contact with the wetting layer,
enabling the activation of hydrodynamic draining modes. Again, the
crossover to the asymptotic behavior is dependent on the surface field
strength and surface morphology, and can be substantially delayed for
strongly-layered surface structures.

\section*{Acknowledgements}

S.P. is grateful to 
Department of Mathematics, Rutgers University, where this work was
initiated.  Research supported in part by NSF Grant DMR-9813268 and
AFOSR
Grant F49620-98-1-0207.

\newpage
\begin{figure}
\centerline{{\epsfbox{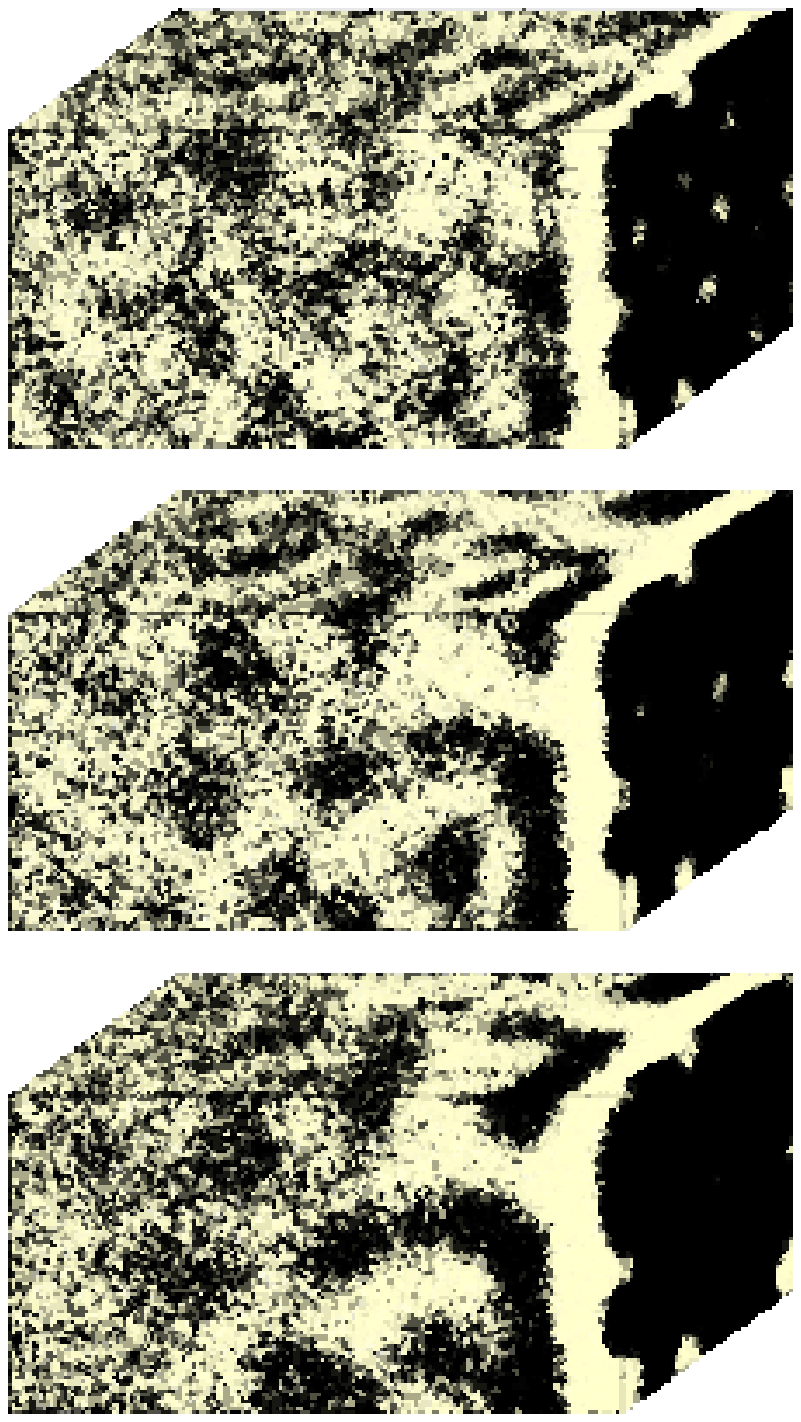}}}
\caption{Snapshots of the evolution after a critical 
quench for a weak-field
case $W_0/W^Y_0=0.67$. The equilibrium surface morphology is partially
wet. The times corresponding to the pictures are (from top to bottom)
$t=60, 120, 180$.}
\end{figure}

\newpage
\begin{figure}
\centerline{\psfig{file=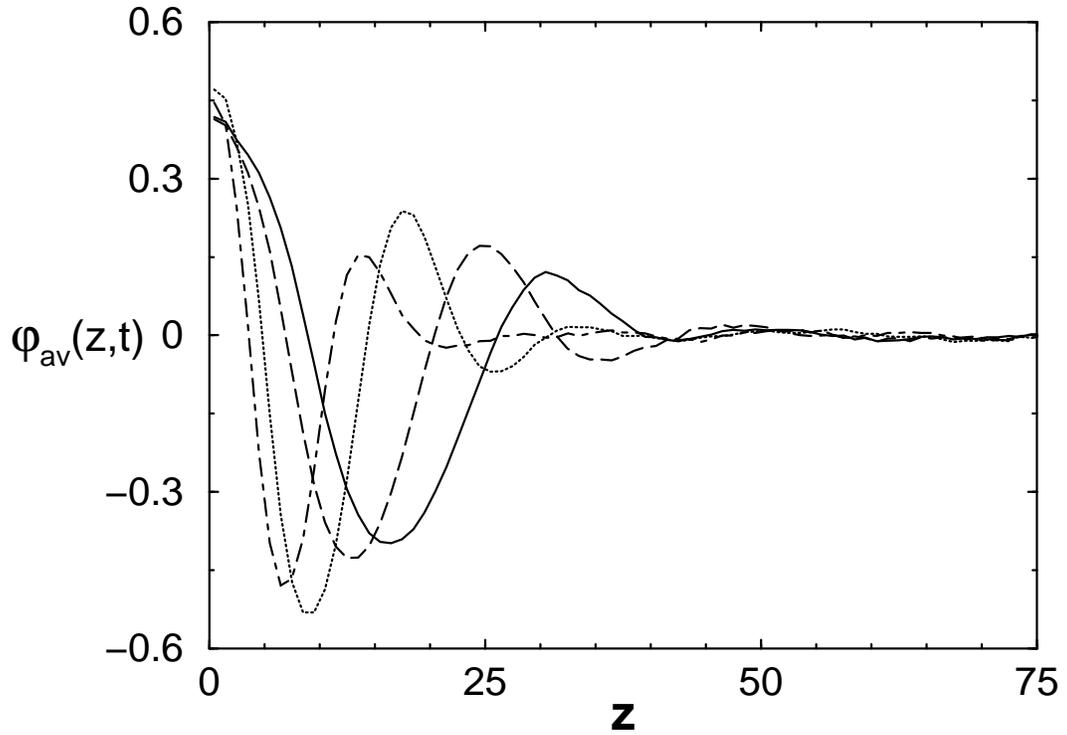,width=5.5truein,angle=-90}}
\caption{Laterally-averaged order parameter profiles, 
$\phi_{av} (z,t)$, as a function of $z$, the distance from the surface.
Parameter values are the same as those for the evolution depicted in
Fig. 1. The evolution times are $t=50$ (dot-dashed), 
$100$ (dotted), $200$ (dashed) and $275$ (solid line).}
\end{figure}

\newpage
\begin{figure}
\centerline{\psfig{file=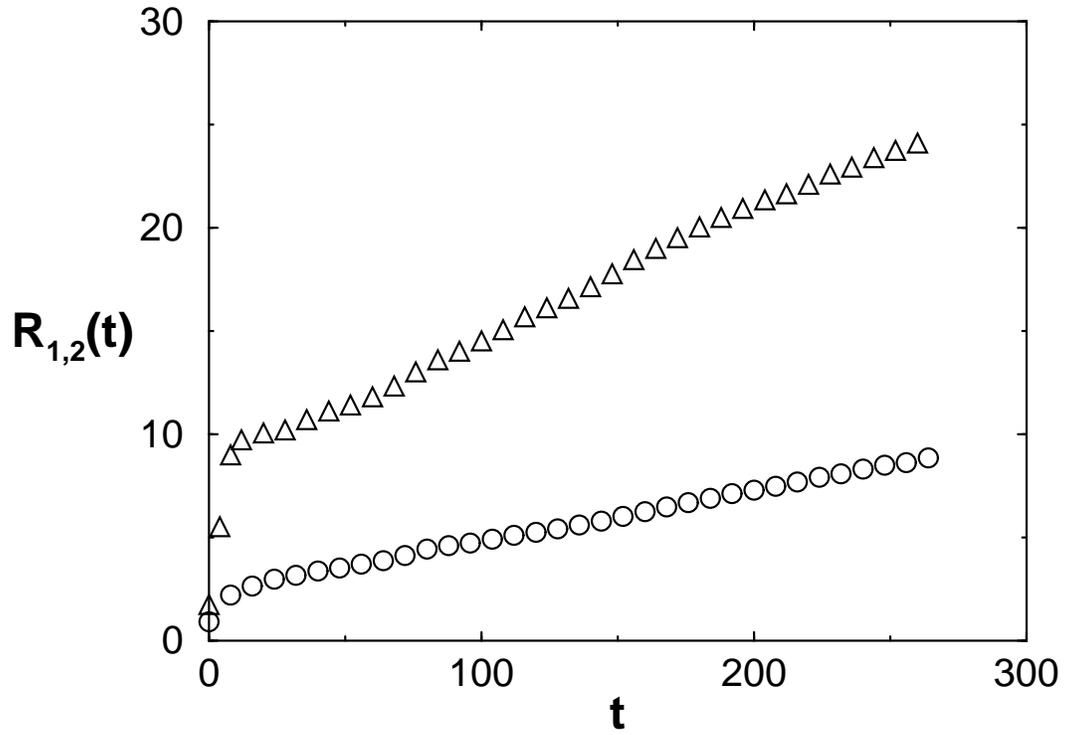,width=5.5truein,angle=-90}}
\caption{Time-dependence of first zero $R_1(t)$ (circles); and second
zero $R_2(t)$ (triangles), of the laterally-averaged 
profiles shown in Fig. 2.}
\end{figure}

\newpage
\begin{figure}
\centerline{{\epsfbox{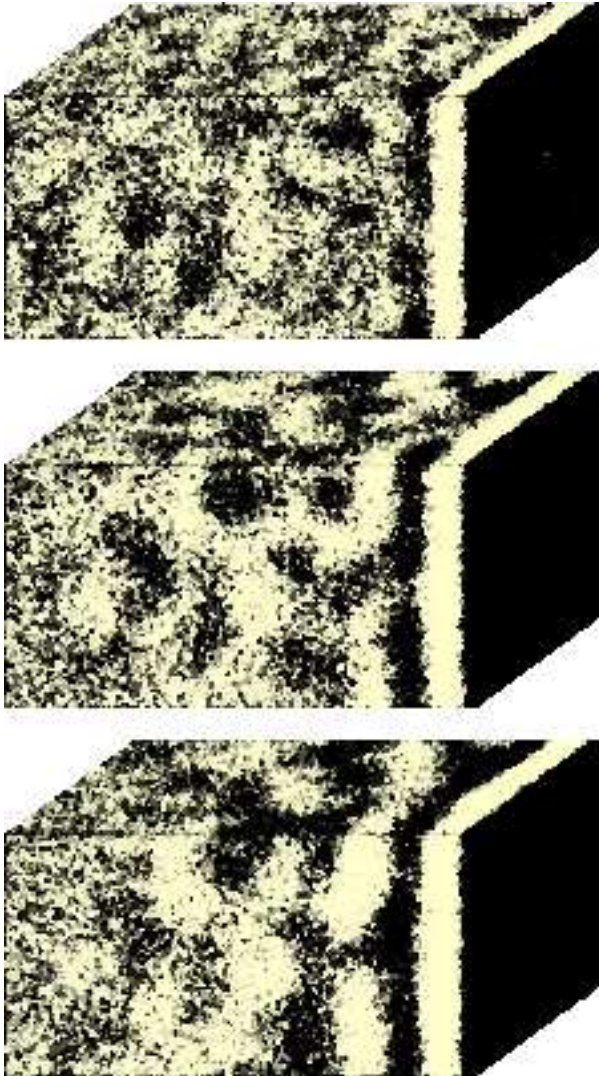}}}
\caption{Analogous to Fig. 1, but for a (very) strong-field case
$W_0/W^Y_0=4$, where the surface is completely wetted by the
preferred component in equilibrium.}
\end{figure}

\newpage
\begin{figure}
\centerline{\psfig{file=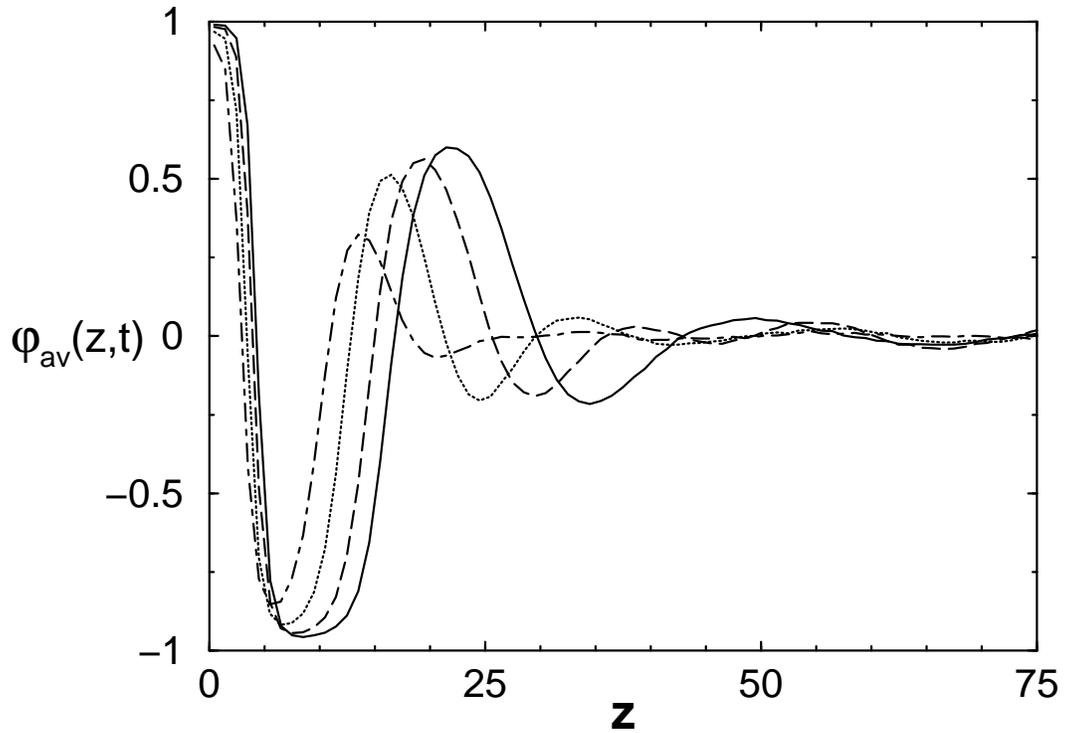,width=5.5truein,angle=-90}}
\caption{Analogous to Fig. 2, but for the evolution depicted in Fig. 4
for the strong-field case. The profiles are shown at times
$t=50$ (dot-dashed), $100$ (dotted), $200$ (dashed) and $275$ (solid
line).}
\end{figure}

\newpage
\begin{figure}
\centerline{\psfig{file=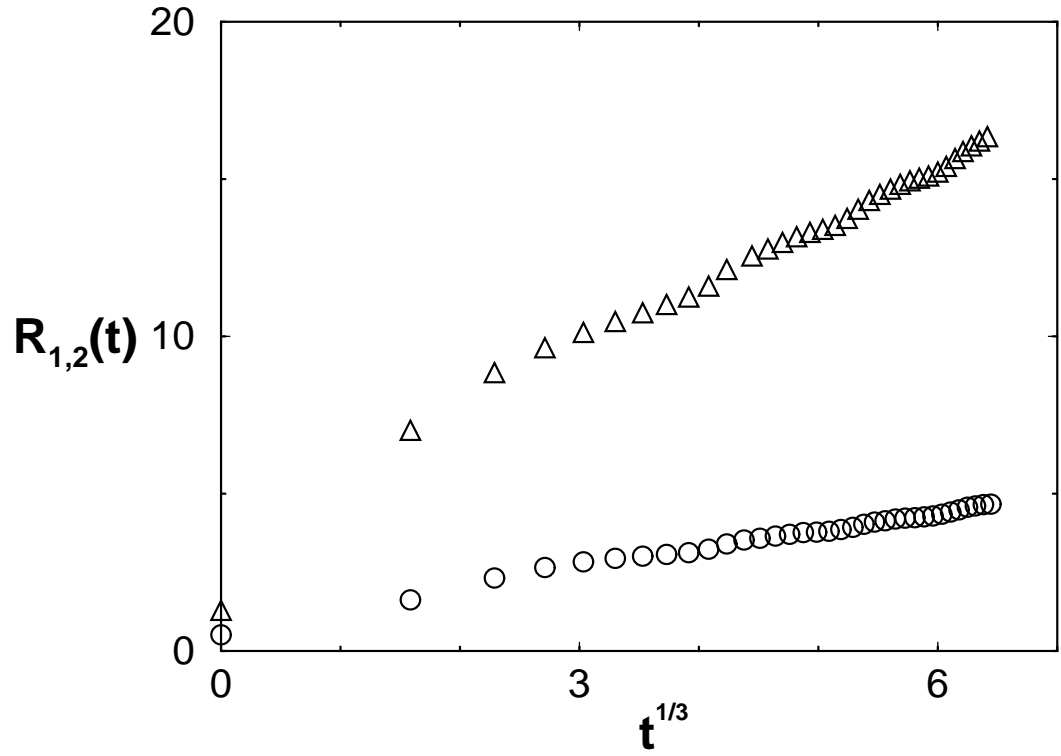,width=5.5truein,angle=-90}}
\caption{Plot of first zero $R_1(t)$ vs. $t^{1/3}$ (circles); and 
second zero $R_2(t)$ vs. $t^{1/3}$ (triangles), of the
laterally-averaged 
profiles shown in Fig. 5.}
\end{figure}

\newpage
\begin{figure}
\centerline{\psfig{file=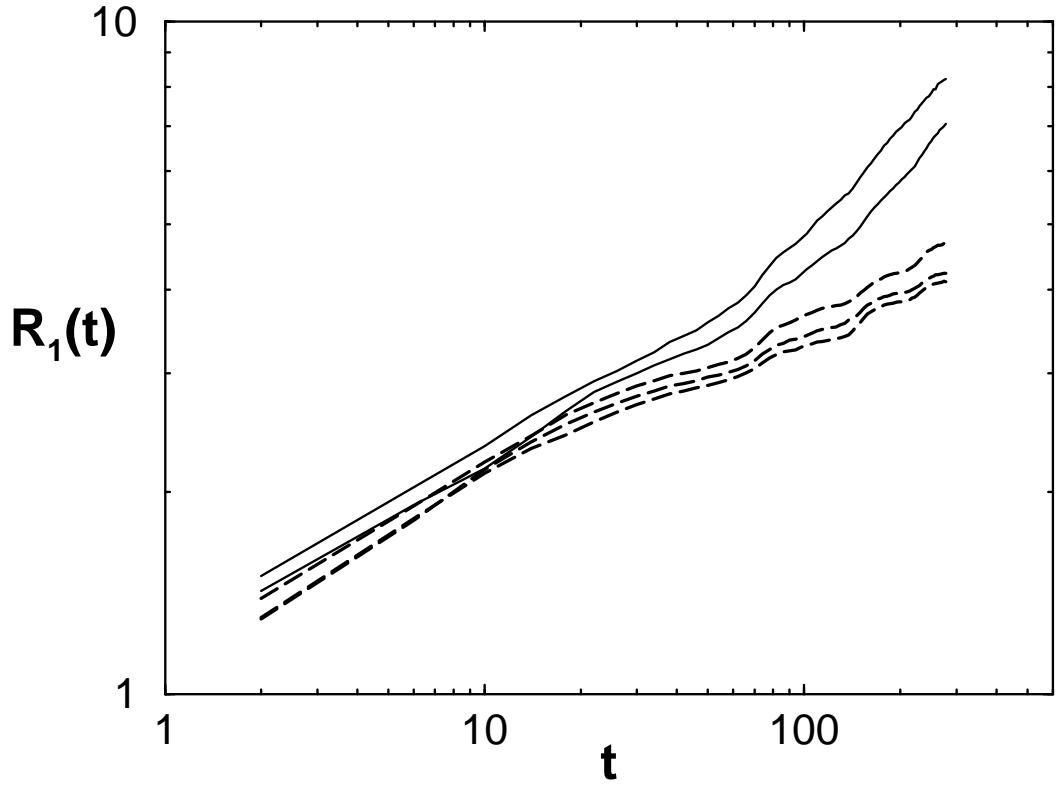,width=5.5truein,angle=-90}}
\caption{Time evolution of the first zero, $R_1$, of the
laterally-averaged profile for all 
cases studied: from top, $W_0/W^Y_0=0.67, 1.33, 2.67, 4.0$ and $5.33$.}
\end{figure}

\end{document}